\documentclass[12pt,a4paper,final]{iopart}

\usepackage{iopams}
\usepackage{graphicx}
\usepackage[breaklinks=true,colorlinks=true,linkcolor=blue,urlcolor=blue,citecolor=blue]{hyperref}

\begin{document}

\title{Multipath interference from large trapped ion chains}

\author{P. Ob\v{s}il$^{1}$, A. Le\v{s}und\'{a}k$^{2}$, T. Pham$^{2}$, G. Araneda$^{3}$, M. \v{C}\'{i}\v{z}ek$^{2}$, O. \v{C}\'{i}p$^{2}$, R. Filip$^1$ and L. Slodi\v{c}ka$^{1}$}
\ead{slodicka@optics.upol.cz}

\address{$^1$ Department of Optics, Palack\'{y} University, 17. listopadu 12, 771 46 Olomouc, Czech Republic \\
$^2$ Institute of Scientific Instruments of the Czech Academy of Sciences, Kr\'{a}lovopolsk\'{a} 147, 612 64 Brno, Czech Republic\\
$^3$ Institut f\"{u}r Experimentalphysik, Universit\"{a}t Innsbruck, Technikerstra\ss e 25, 6020 Innsbruck, Austria}

\begin{abstract}
The demonstration of optical multipath interference from a large number of quantum emitters is essential for the realization of many paradigmatic experiments in quantum optics. However, such interference remains still unexplored as it crucially depends on the sub-wavelength positioning accuracy and stability of all emitters. We present the observation of controlled interference of light scattered from strings of up to 53~trapped ions. The light scattered from ions localized in a harmonic trapping potential is collected along the ion crystal symmetry axis, which guarantees the spatial indistinguishability and allows for an efficient scaling of the contributing ion number. We achieve the preservation of the coherence of scattered light observable for all the measured string sizes and nearly-optimal enhancement of phase sensitivity. The presented results will enable realization and control of directional photon emission, direct detection of enhanced quadrature squeezing of atomic resonance fluorescence, or optical generation of genuine multi-partite entanglement of atoms.
\end{abstract}

\section{Introduction}

In atomic physics, optical coherence constitutes an exquisite tool for testing fundamental theories and its control has provided many dramatic advancements in practical applications of controlled atomic systems~\cite{cohen2011advances,inguscio2013atomic}. Optical coherence can be accessed by the precise knowledge of the interaction of light with the traversed environment which, at its most fundamental level, can be studied down to individual atoms. However, the coherence of light scattered by individual atoms has proven to be extremely sensitive to several intrinsic and environmental effects which affect the atomic positioning or phase of the emitted light in essentially untraceable way~\cite{predojevic2015engineering}. Therefore, although the coherent light scattering from ensembles of individual atoms corresponds to a crucial tool for investigation of many fundamental phenomena in quantum optics and could provide a viable approach for scaling up of some of its exquisite applications~\cite{lehmberg1970radiation, porras2008collective, skornia2001nonclassical, sutherland2016collective, scully2009collective, kauten2017obtaining, ludlow2015optical,wehner2018quantum}, it has been almost exclusively observed in systems consisting of very few atoms~\cite{eichmann1993young, itano1998complementarity, eschner2001light, wolf2016visibility, slodicka2012interferometric,neuzner2016interference}. Control of optical coherence in individually addressable atomic ensembles would contribute to understanding of the bare light interference in multi-path arrangements, where the effect of various global or localized experimental imperfections, including phase jitters or positioning errors, can be hard to predict reliably for large number of paths. Preservation of optical coherence upon scattering from individual atoms would allow for adaptation of well developed methods from crystallography~\cite{patterson1935direct, ladd2013structure} for studies of position and nonclassical motion in atomic experiments~\cite{shahmoon2018quantum} and, conversely, enable emulation of crystallographic phenomena in controllable atomic ensembles. In addition, scaling up the number of atoms without significant loss of coherence would directly imply the possibility of optical generation of multi-partite entangled states~\cite{cabrillo1999creation, slodivcka2013atom, thiel2007generation, maser2010versatile, ficek2002entangled}, controllable design of collective atomic emission~\cite{vogel1985squeezing, devoe1996observation, jin2010squeezing, mewton2007radiative, clemens2003collective}, or realization of a feasible alternative for enhancement of absorption and collection of light in few-atom systems~\cite{casabone2015enhanced, oppel2014directional, clemens2003collective, mewton2007radiative}.

The broad application prospects have stimulated a number of pioneering experiments, where the sub-wavelength localization of contributing atoms has become evident in the observable interference of scattered light~\cite{birkl1995bragg, Weitenberg2011coherent, weidemuller1995bragg, slama2005phase, neuzner2016interference, raithel1997cooling, schilke2011photonic, sorensen2016coherent, eichmann1993young, itano1998complementarity, eschner2001light, wolf2016visibility, slodicka2012interferometric}. They have been technologically approached mainly from two complementary directions: one employing the convenient scalability of number of neutral atoms confined in individual optical lattice sites~\cite{birkl1995bragg, Weitenberg2011coherent, weidemuller1995bragg, slama2005phase, neuzner2016interference, raithel1997cooling, schilke2011photonic, sorensen2016coherent} and other exploiting the natural sub-wavelength localization and individual addressability of atomic ions stored in Paul traps~\cite{eichmann1993young, itano1998complementarity, eschner2001light, wolf2016visibility, slodicka2012interferometric}.
These manifestations of coherent scattering from ordered atomic ensembles have even enabled several proof-of-principle demonstrations of sensing~\cite{wolf2016visibility, slodicka2012interferometric, araneda2018interference, raithel1997cooling, weidemuller1998local}. However, the demonstration of the envisioned system corresponding to ensembles of individual coherent atomic scatterers capable of the long-term sub-wavelength spatial localization, with the possibility of their individual control and essential scalability with respect to number of contributing atoms, has long remained a desired milestone.

We report on the observation of multi-path interference of light scattered from large strings of trapped atomic ions. Importantly, the phase of the interference is controllable and all ions in the string contribute to the detected light with near-equal amplitude.
The implemented detection along the symmetry axis of the linear trapping potential, together with the precise control of motion and excitation parameters of contributing ions, allow for scaling up the ion number simultaneously contributing to coherent light scattering without considerable effect on their relative fluorescence collection efficiency. The spatial coherence of the detected light is guaranteed by its coupling to a single-mode optical fiber and the temporal and spectral coherence is maximized by optimization of the laser cooling and laser excitation parameters. Simultaneous achievement of all these experimental conditions was crucial for demonstrated preservation of interference visibility for up to 53~ions and narrowing the width of interference peak to less than one third when compared to the two-ion case.

\section{Experimental scheme}

\begin{figure}[h]
\begin{center}
\includegraphics[width=0.75\columnwidth]{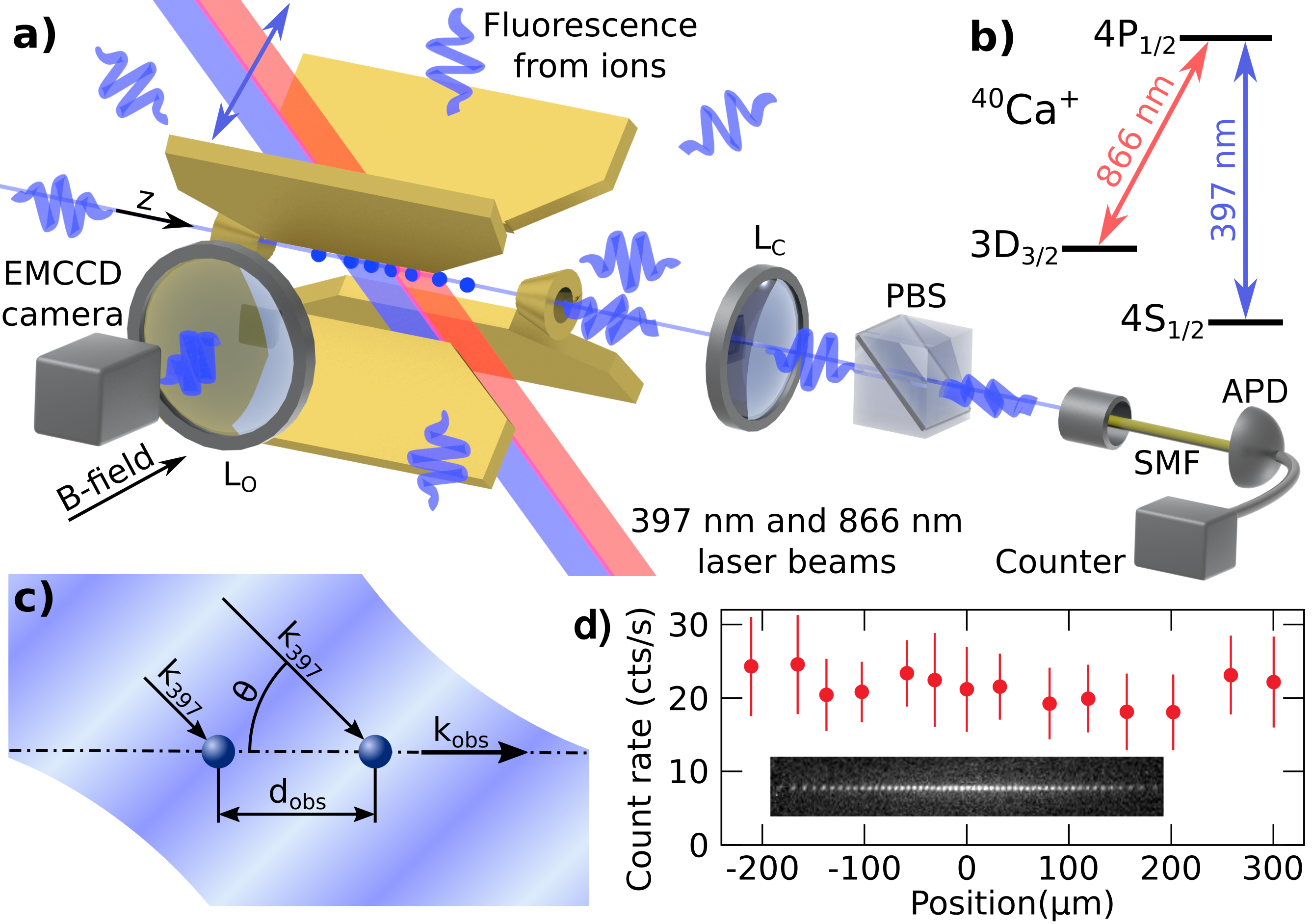}
\caption{a) A simplified scheme of the employed experimental setup. Ions confined in the linear Paul trap scatter the light from the exciting 397~nm laser beam and its phase interference is observed along the trap symmetry axis through a small aperture in the tip electrode. The light is collected by the coupling lens $L_c$ and its polarization is filtered using polarization beam splitter (PBS) to maximize the portion of elastically scattered photons in the detected signal. After coupling into the single-mode fiber (SMF), the scattered light is detected by the single-photon avalanche photodiode (APD). The phase of the interference of scattered light is tuned by adjustment of the applied axial potential strength through voltages $U_{\rm tip}$. The auxiliary observation channel used for independent estimation of the mutual ion distance and ion counting is realized along the direction of the applied magnetic field $\vec{B}$. b) A simplified energy level scheme of $^{40}$Ca$^+$ with the respective transition wavelengths. c) A scheme of the employed geometric arrangement and corresponding interfering photon paths. The measured data in d) show the dependence of the detected photon rate from single trapped ion as a function of its axial spatial position $z$. The inset shows the 53-ion crystal captured at the lowest employed tip voltage of $U_{\rm tip}=5.3$\,V to illustrate the spatial extent of measured ion strings.}
\label{fig:block_scheme}
\end{center}
\end{figure}

The Fig.~\ref{fig:block_scheme}-a) shows schematically the experimental arrangement. It allows for the observation of scattered light from the controllable atomic system with a well-defined number of ions and with the fundamental possibility of their individual addressing. The experimental setup comprises a linear Paul trap with four blade-like electrodes for spatial confinement of ions in the radial direction defined as the direction perpendicular to the trap symmetry axis. The ultra-high-vacuum surrounding the Paul trap has estimated pressure on the $10^{-12}$~mBar level, which enables long storage and measurement times with large ensembles of ions. The observation of the ions on an EMCCD camera allows for the spatial resolution of each individual ion and it is used for direct determination of number of ions in the string, its length, and overall emission intensity. In the axial trapping direction, ions are confined by application of dc voltages $U_{\rm tip}$ on the two cone-shaped tip electrodes. The employed tip electrodes have small apertures which are typically utilized for quasi-homogeneous excitation of the ion crystals~\cite{schindler2013quantum, haffner2008quantum}. Conversely, we use these apertures for the collection of light scattered by the ion strings. After coupling the fluorescence to a single-mode optical fiber and optimization of the collection optics, the presented configuration corresponds to a robust detection scheme with several crucial properties. First, for the relatively small solid angles offered by the apertures,
the light scattered from ions aligned along the optical axis of the collection optics is detected with approximately the same efficiency as far as they lay within the Rayleigh range of the detection mode. Second, the axial detection configuration also allows for convenient bulk detection of the scattered light interference, because the angular gradient of the interference pattern of the collectively scattered light is expected to be the smallest close to the direction of the ion crystal axis~\cite{mewton2007radiative}. Therefore, the presented spatial configuration allows for maximizing the optical access with minimal effect on the observed interference visibility.

The $^{40}$Ca${^+}$ ions are trapped with parameters chosen to achieve linear ion strings along the axial trapping direction. The radial secular frequencies are set to $f_x \approx f_y \approx 1.66$ MHz, while the axial potential is tuned in the frequency range $f_z \approx 60$~kHz$ - 1044$~kHz, corresponding to $U_{\rm tip}=4$~V and $U_{\rm tip}=900$~V, respectively. The micromotion has been carefully compensated for all employed $U_{\rm tip}$ settings.  The compensation has been carried out with single ion by detection of the temporal correlation between the trapping radio frequency signal and detected fluorescence using the 397~nm excitation beam~\cite{berkeland1998minimization}. The ions are Doppler cooled using a 397~nm laser red-detuned from the 4S$_{1/2}\leftrightarrow$ 4P$_{1/2}$ transition, see Fig.~\ref{fig:block_scheme}-b) 
for the relevant energy level scheme. The population of the 3D$_{3/2}$ manifold coming from the 4P$_{1/2}\rightarrow {\rm 3D}_{3/2}$ decay is reshuffled to the Doppler cooling transition by 866~nm laser light. The degeneracy of Zeeman states is lifted by the application of a static magnetic field with magnitude $|\vec{B}|=6.1$~Gauss along the direction of observation on the EMCCD camera. The fluorescence in the axial trapping direction is collected using a single plano-convex lens $L_{\rm c}$ with  focal length $f=150$\,mm positioned about $150\pm2$~mm from the trap center and outside of the vacuum chamber. The collected light passes a polarization beam-splitter (PBS) set for transmission of the light scattered on the $\sigma^+$ and $\sigma^-$ atomic transitions in order to maximize the portion of the observed elastic scattering for the given excitation laser direction and polarization. The fluorescence is coupled to a single-mode optical fiber (SMF) for the precise definition of its spatial mode and detected using the single-photon avalanche photodiode (APD) with specified quantum efficiency of $77$\%. Together with the measured losses of optical components and estimated effective solid angle fraction of $1.3\times 10^{-5}$, the evaluated overall detection efficiency $\eta_{\rm e}=3.7\times 10^{-4}$\,\% of the light emitted from ion positioned in the center of the trap corresponds very well to the measured $\eta_{\rm m}=(3.6 \pm 0.4) \times 10^{-4}$\,\%. The detection efficiency has been estimated by recording the 397\,nm fluorescence after the excitation of spontaneous Raman transition 3D$_{3/2}\rightarrow {\rm 4P}_{1/2} \rightarrow {\rm 4S}_{1/2}$ from an optically pre-pumped 3D$_{3/2}$ manifold~\cite{higginbottom2016pure,obvsil2017nonclassical}. In the presented experimental setup, this number is mainly limited by the effective solid angle fraction allowed by the apertures in the tip electrodes, however, these could be easily engineered to much larger diameters while still allowing to efficiently capture the light from tens to hundreds of trapped ions. We have estimated the width of the detection optical mode at the center of the trap by measuring the efficiency of photon detection as a function of the spatial displacement of an ion in the radial trapping direction. The Gaussian fit of the measured dependence results in detection mode waist of $w_0=17\,\mu$m corresponding to a Rayleigh length of $z_R=2.3$\,mm.

The measured dependence of the detection probability of 397\,nm fluorescence as a function of the axial position of a single trapped Ca${^+}$ ion is shown in Fig.~\ref{fig:block_scheme}-d). It demonstrates the near-constant collection efficiency within the measurement uncertainties given by the shot noise and by the repeatability of the laser excitation parameters setting and within the total axial spatial position range of more than 500~$\mu$m.
The largest crystal employed in the presented measurements corresponding to 53\,ions has at its stretched setting corresponding to applied axial potential $U_{\rm tip}$=5.3~V estimated spatial length of 367$\mu$m.
To further confirm the equality of contribution from different ions to detected light field, we have precisely measured the scaling of the detected photon rate in the pulsed generation regime for 19, 39 and 64 ions in close analogy with measurements presented in~\cite{higginbottom2016pure, obvsil2017nonclassical}. We have observed linear dependence of the overall detected count rate for up to 64~ions and largest crystal length of about 340\,$\mu$m. These measurements clearly show the unprecedented uniformity of the fluorescence coupling efficiency into the single spatial mode from long ion strings in the presented setup. Similar close-to-linear scaling is expected for even longer ion strings provided that they would be still well within the Rayleigh range of the observation light mode.

\section{Observation of coherent light scattering}

\subsection{Two ions}

The two-ion light scattering measurements can give us important information about the relative contributions of particular decoherence mechanisms in our setup and estimation of the expectable limit on the visibility for larger numbers of ions. As can be seen in the schematic drawing in Fig.~\ref{fig:block_scheme}-c), the particular two-ion interference path arrangement can be understood as atomic version of a classical Michelson interferometer. The 397~nm laser light propagating with an angle $\theta\sim45^\circ$ with respect to the ion crystal axis can be, with a small probability, reflected off the two ions. The relative distance between two ions in the observation direction $d_{\rm obs}$ projected on the excitation laser beam direction gives $d_{\rm exc}=d_{\rm obs}\cos\theta$, which results in the overall relative phase difference $\delta\varphi=d_{\rm obs} (1-\cos\theta) \omega/c$, where $\omega$ is the angular frequency of the scattered light. The distance $d_{\rm obs}$ can be tuned by the controlling the voltage $U_{\rm tip}$ and its setting can be compared with the theoretically expected values~\cite{james1998quantum} by independent measurement of $d_{\rm obs}$ on the EMCCD camera.

\begin{figure}[htp]
\begin{center}
\includegraphics[width=0.7\columnwidth]{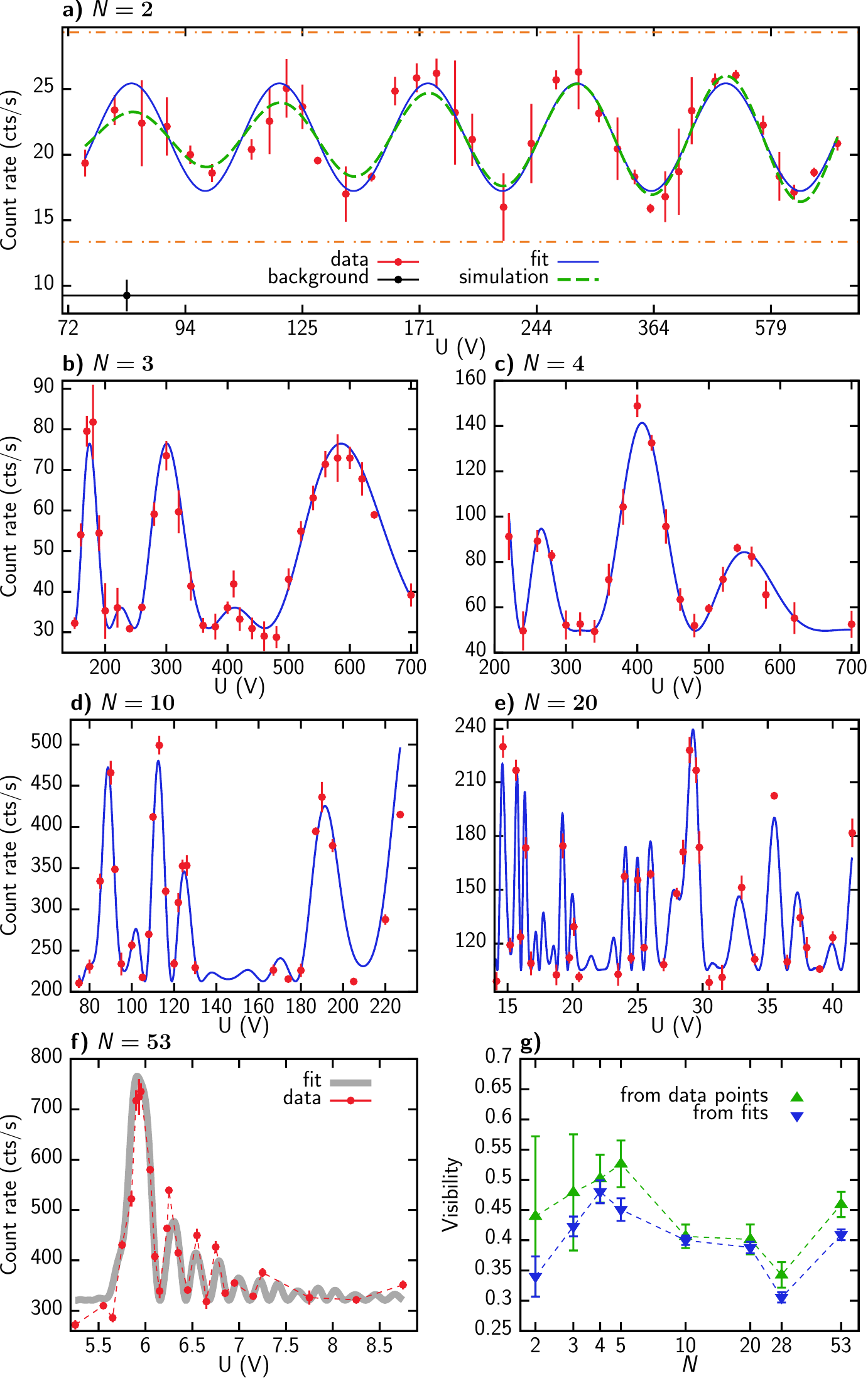}
\caption{Results of the measurements of phase interference of light scattered from ion crystals containing from 2 to 53~ions. The red points correspond to the measured count rates and blue solid lines show the theoretical model based on the numerical calculation of the mean relative ion positions for the given axial voltages $U_{\rm tip}$, with amplitude and offset fitted to the data. All error bars correspond to statistically evaluated single standard deviation. Fig.~a) shows the interference measured with 2~ions together with the full theoretical simulation taking into account the main known decoherence mechanisms (green dashed line). The measured background counts (black data point) are same for all axial voltage settings. The orange dash-dotted lines depict the estimated interference limits given by the portion of inelastically scattered light. Fig.~b)-e) show samples of the interference patterns from 3 to 20~ions, respectively. Note that for more than 3~ions the mutual ion spacing is not uniform, so the expected interference patterns are not those of a regular grid. The data in~f) correspond to observations of interference with $n=53$ ions, where the Gaussian spatial beam intensity profile had to be taken into an account in order to fit the observed interference pattern plausibly (gray line). The measured visibility scaling with the ion number $n$ estimated from the extremal data points and from fits is plotted in g).}
\label{fig:interference}
\end{center}
\end{figure}

The data in Fig.~\ref{fig:interference}-a) show measured count rates of photons scattered from two-ion crystal as a function of tip-electrode voltage, i.e., of the ion distance, with 20\,s overall integration time. The error bars were evaluated statistically from 5\,s~long data samples. The observed oscillation behavior can be reproduced by simple calculation of the observed total light intensity dependence on the relative phase $I_{\rm tot} = 2 I_1 (1+|g^{(1)}(\tau)| \cos\delta\varphi)$ shown as a solid blue curve. Here $I_1$ is the average light intensity from a single ion in the absence of the second ion and the $g^{(1)}(\tau)$ is the degree of first-order coherence. $|g^{(1)}(\tau)|$ in Fig.~\ref{fig:interference}-a) has been scaled to find the best fit to the measured data, which is then used for estimation of the raw interference visibility $V=(I_{\rm tot}^{\rm max}-I_{\rm tot}^{\rm min})/(I_{\rm tot}^{\rm max}+I_{\rm tot}^{\rm min})=|g^{(1)}(\tau)|=19$\,\%. Here $I_{\rm tot}^{\rm max}$ and $I_{\rm tot}^{\rm min}$ are the maximum and minimum count rates from the fitted interference pattern, respectively. The estimated visibility after subtraction of spurious background counts of ($9.3\pm1.2$)~cts/s is $V=34$\,\%, which is comparable to other observations of the phase interference with trapped ions after Doppler cooling~\cite{slodicka2012interferometric, wolf2016visibility, eschner2001light}. It is apparent, that the observed background counts, in our case almost solely caused by detector dark counts, can contribute considerably to the detected count rate in the few-ion case for the given small collection efficiency. The phase factor $\delta\varphi$ in the presented fits has been set by independent estimation of the two-ion distance $d_{\rm obs}$ from calibrated axial trap frequencies $f_z$ corresponding to axial tip voltages $U_{\rm tip}$. The angle $\theta$ between the excitation and observation directions $\theta=(45.19\pm0.04)^\circ$ has been found by the best fit to the measured data, in good agreement with the independently measured value of $\theta_m=(45.24\pm0.06)^\circ$.

The decrease of interference contrast in experiments concerned with scattering of the coherent light from trapped ion crystals is mostly caused by inelastic light scattering and motional dephasing mechanisms~\cite{eschner2001light, slodicka2012interferometric, wolf2016visibility, wong1997interference}. For the two-ion case, we have estimated the ratio of elastic to inelastic scattering by the independent analysis of the laser excitation parameters using dark-resonance spectroscopy and a corresponding model based on eight-level optical Bloch equations. The effective saturation parameter of the 397\,nm laser on the 4S$_{1/2}\leftrightarrow {\rm 4P}_{1/2}$ transition has been found to be $s_{397}=0.51$, which, according to numerical simulations, accounts for the reduction of the observable visibility by a factor of~0.66. The jitter in the relative position of the two ions due to their secular motion can be taken into account by including the Gaussian position uncertainties corresponding to the thermal population of the radial zig-zag and axial stretch modes with motional frequencies and mean phonon numbers measured separately for the same trapping and laser-cooling conditions~\cite{slodicka2012interferometric, wolf2016visibility}. The simulation taking into account both reduction mechanisms is shown as the green dashed line in Fig.~\ref{fig:interference}-a). It reproduces the measured interference pattern without any free fitting parameter, what again underlines the high degree of control of both motional and internal excitation parameters in the presented ion-trapping apparatus. The clear visibility decrease for lower axial confinements is explained by the corresponding increase in the relative position uncertainty.

\subsection{Many ions}

Figs.~\ref{fig:interference}-b) to e) present the measured interference patterns for 3, 4, 10, and 20~ions, respectively. Since the trapping potential produced by the Paul trap is approximately harmonic, the separation between ions for more than 3~ions is nonuniform. The nonuniform distribution makes interference fringes nonperiodic, however, it does not break interference effect itself.
The observed interference patterns clearly demonstrate the ability of large ion strings to scatter light in a partially coherent way.
The blue curves in Figs.~\ref{fig:interference}-b) to e) correspond to simulations, which assume equal coherent contributions from all ions with the phase factor calculated according to the numerically estimated mean ion positions given by $U_{\rm tip}$~\cite{james1998quantum}.
The simulations require only the knowledge of the tip voltages $U_{\rm tip}$ and the angle between the excitation and observation direction $\theta$. The precise position of each ion in the trap along the axial trapping direction can be evaluated using the relative coefficients in equation (5) from ref.~\cite{james1998quantum}. The axial center-of-mass~(COM) motional frequency has been found experimentally on single trapped ion and adjusted for the given particular number of ions using the same equation. The relative position of individual ions allows us to
calculate the relative phase shift $\Delta \phi$ relevant for the observed interference pattern. The observed count rate is then, in the ideal case of 100\,\% interference visibility and fully coherent light scattering process, proportional to light intensity
\begin{equation}
I_{\rm coh}=\left| \sum_{j=1}^N A_j e^{i \Delta \phi_j}\right|^2,
\end{equation}
where $A_j$ is the amplitude of light scattered by \emph{j}-th ion in the observation direction. $A_j$ has been set to the unity value for all contributing ions from the string for all measured ion strings containing up to 20 ions. The obtained ideal interference pattern as a function of tip voltages $U_{\rm tip}$ has been then scaled to best fit of the measured count rate using
\begin{equation}
I_{\rm obs}=I_{\rm incoh}+\kappa\cdot I_{\rm coh},
\end{equation}
where we fit parameters $I_{\rm incoh}$ and $\kappa$ correspond to the incoherent light offset including the detected background counts and relative coherent contribution, respectively.
The visibilities of simulated interference fringes provide an estimate of the local interference visibility $V$ in analogy with the two-ion case. The simulations are in very good agreement with the observed interference patterns, which suggests that the spatial distributions of ions in the crystal correspond to the ideal positions given solely by the axial trapping potential and the mutual Coulomb repulsion. In addition, this agreement points to the near-equality of the coherent contribution to the detected signal from all individual ions present in the crystal, see Appendix for quantitative comparisons. The precisely measured background corresponding dominantly to detector dark counts has been subtracted in order to unify the presentation with the two ion case, however, it presents a much less significant limit for these large ion strings.

The measured data in Fig.~\ref{fig:interference}-f) show the observed interference pattern for the string containing 53~ions. In the the corresponding simulation, the amplitude of the light coherently scattered by ions at different axial spatial positions $z$ is not constant and includes the Gaussian spatial modulation of the exciting 397~nm beam intensity,
\begin{equation}
A_j(z)=A_j(0)\frac{1}{\sqrt{2 \pi} \sigma_z}\exp(-z^2/(\sigma_z^2)).
\end{equation}
where $\sigma_z=115\,\mu$m corresponds to the measured width of 397~nm laser beam intensity profile projected on the axial direction. The excitation beam width presents technical limit of this particular optical setup, however, it could be relatively easily overcome by the employment of the elliptical laser beams or smaller scattering angle. To support our conclusions about the source of modulation of the detected count rate in the presented data, we have also checked the total intensity of the ion string in the observation mode of the EMCCD camera, where it always remained constant within the measured ranges of $U_{\rm tip}$. We note, that for very low axial trapping potentials, a slight positive difference between $U_{\rm tip}$ values in simulation and in the experiment has been required for plausible reconstruction of the observed interference patterns. The magnitude of this deviation is several tens of mV at several volts of applied $U_{\rm tip}$, which could be likely explained by the residual stray fields or patch potentials seen by the ions~\cite{harter2014long, brownnutt2015ion}.

The observed scaling of the interference visibilities is summarized in Fig.~\ref{fig:interference}-g). The visibilities estimated from extremal data points in the locally observed interference patterns and from the fitted numerical model are shown as green triangles and blue squares, respectively. Remarkably, they remain within the range 0.34 to 0.53 for all measured ion numbers. The observed dependence of the interference visibility on the number of ions is dominantly given by two competing phenomena. First, the maximum intensity corresponding to a fully constructive interference tends to increase with a number of scatterers as $N^2\times I_{\rm coh}+N\times I_{\rm incoh}$, while the given portion of incoherent scattering corresponding to an interference minimum sums up linearly as $N\times I_{\rm incoh}$. This results in the visibility increase for a small number of ions with a finite incoherent scattering contribution. However, the non-equidistant positions of scatterers with their relative positions given by the bare Coulomb repulsion in the harmonic potential of a Paul trap result in an inability of observation of the global interference maxima within the accessible potential strengths for more than 4 ions in our setup. Therefore, for $N>4$ the two effects compete and the visibility is given by the particular convenience of the achievable potential width interval for the given ion number. In addition, we note, that the laser excitation and Doppler cooling parameters were individually optimized for achieving maximal visibility values for each ion number, because we were not able to find satisfactory global settings working for all ion numbers. In addition, the laser excitation and Doppler cooling parameters were individually optimized for achieving maximal visibility values for each ion number, because we were not able to find satisfactory global settings working for all ion numbers. Due to the large parameter space, these optimizations likely contribute to observed dependence in the measured visibilities. Similarly to the two-ion case, the observed contrast is dominantly limited by the probability of elastic scattering, which could be substantially enhanced by operating in pulsed excitation regime allowing for the separation of cooling and probing processes.

\subsection{Evaluation of measured phase resolution and maximal intensity}
\label{subsec:characteristics}

\begin{figure}[h!]
\begin{center}
\includegraphics[width=0.6\columnwidth]{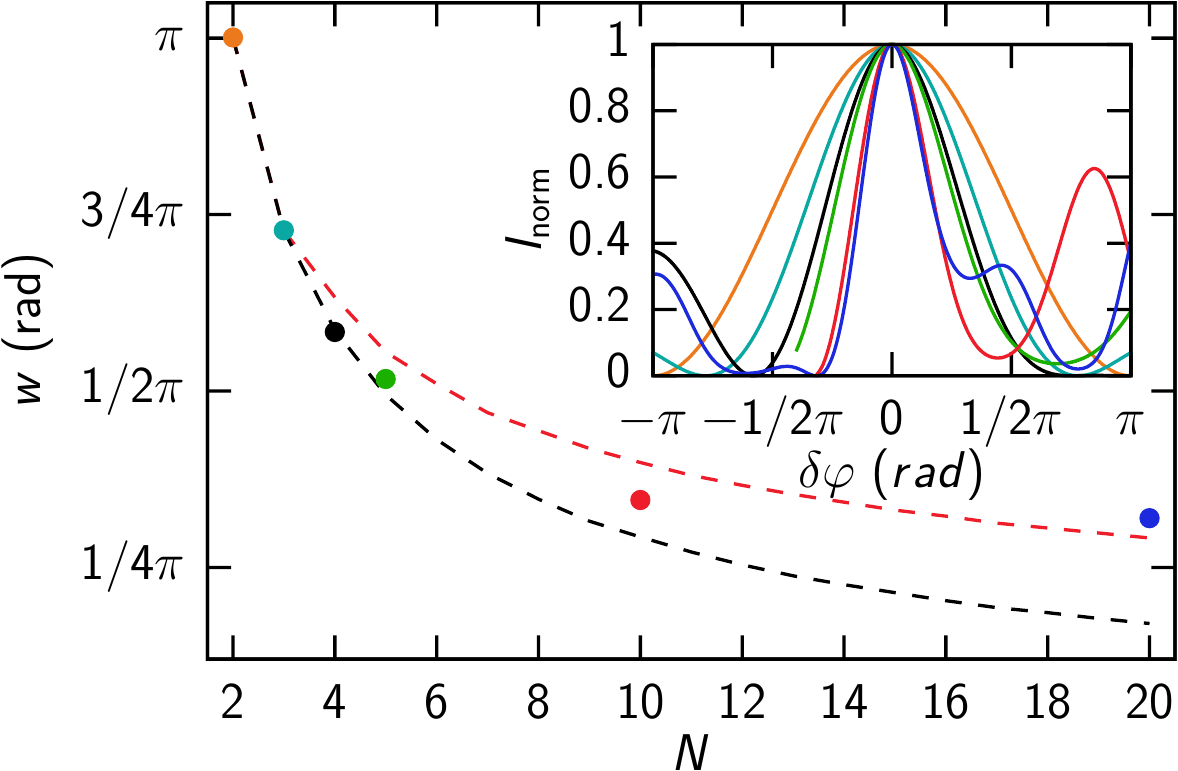}
\caption{The analysis of phase sensitivity of multi-ion interference. The full width at half maximum widths (FWHM) \emph{w} of observed maximal interference peaks for measured strings containing $N$ ions are evaluated from the simulations fitting the experimental data. The inset shows the simulated interference maxima from which the widths were estimated. The colours of the data points in the main graph are set to match the colours of the corresponding interference peak shown in the inset. The spatial dimension on the horizontal axis is normalized to the obtained distance between two ions for a given voltage $U_{\rm tip}$, which unifies the length scale for the realized measurements with different ion numbers and different measured axial voltage ranges. The dashed black and red lines correspond to the widths of maximum intensity peaks in the interference patterns for the equidistant ion strings with the distances scaled to the largest and smallest distances between neighbouring ions in the harmonic potential case. The error bars evaluated from the uncertainties in the applied voltages and excitation angle are below the size of displayed data points.}
\label{fig:fwhm}
\end{center}
\end{figure}

In the context of multi-path interference on charged particles in a static harmonic electric potential, it is particularly interesting to analyse the extractable phase resolution. It has been mostly studied in the case of equally spaced reflectors, which, in the limit of their high number, leads to the description of the conventional Fabry-P\'erot resonator. The case studied here is clearly challenged by this ideal case. The phases of the light scattered by the individual ions are less and less likely to add up perfectly constructively for increased number of scatterers and for the small range of scanned $U_{\rm tip}$. On the other hand, in the range of ion numbers corresponding to several tens, the points of significantly high constructive interference should be still found within the practical limits of the trap. The analysis of the scaling of the observed maximal interference peak width observed within the voltage range where ion crystal remained stable is shown in the Fig.~\ref{fig:fwhm}. The interference from many emitters narrows the width of the maximum peaks, similarly as for symmetrical multi-beam interference. The observable maximum peak width corresponding to the change of the phase delay $\delta\varphi$ steadily decreases from the reference value $\pi\,$rad for two ions for all measurements up to 10~ions. This dependence saturates, as the full width at half maximum (FWHM) of the observed highest interference peaks for 10 and 20 ions are $0.35\pi \,$rad and $0.32\pi \,$rad, respectively. This saturation is the direct consequence of the restriction in the applicable voltages, with $U_{\rm tip}$ varying from 5~V to 900~V, which for ion numbers beyond four doesn't cover the full period of the interference pattern. In turn, the observable enhancement of the spatial phase resolution within the measured range of ion numbers signifies the increase of number of coherently contributing scatterers. The achievable phase resolution could be further enhanced by regularization of the atomic spatial pattern using optical pumping~\cite{sorensen2016coherent}. We note that the measurements on more than 20 ions are excluded here due to the limiting spatial extent of the exciting 397~nm laser beam, which effectively causes broadening of the interference pattern.

\begin{figure}[h!]
\begin{center}
\includegraphics[width=0.6\columnwidth]{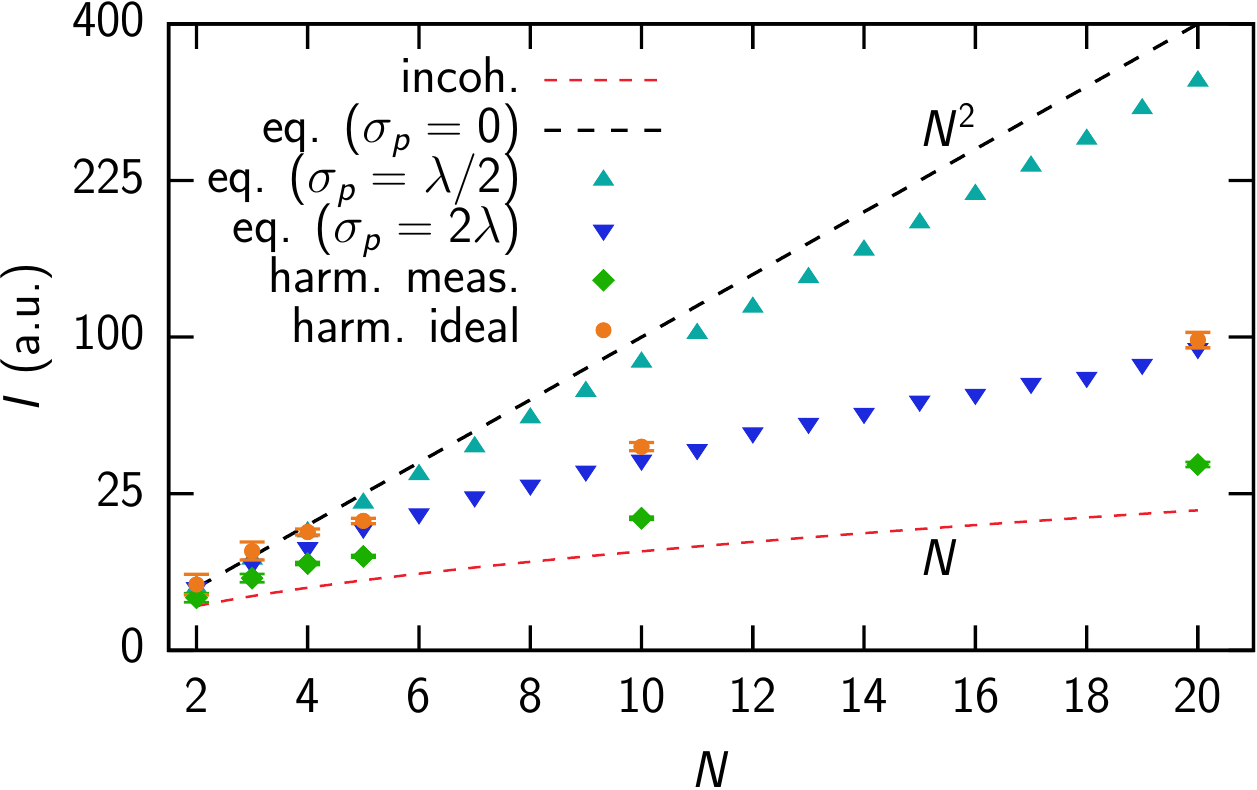}
\caption{Analysis of the maximal observable scattered light intensity in the measured interference patterns. The two limiting cases of fully incoherent scatterers and fully coherent equidistant scatterers mark the relevant intensity scale shown as red and black dashed line, respectively. All the data are normalized to unit intensity per scatterer. Note the square scaling of the vertical graph axis. The normalized measured maximal intensities of light scattered by emitters in the harmonic potential (green diamonds) are always substantially surpassing the fully incoherent case. If the scattering would be considered fully coherent, the measured data would be approaching the limit (orange circles) given by the substraction of the incoherent background. For comparison with the equidistant case, we show the case of fully coherent scattering by quasi-equidistant scatterers with a Gaussian distribution of individual positions around the ideal equidistant case (blue triangles). \emph{Eq.} and \emph{harm.} denote the equal and harmonic spacing case, respectively, and $\sigma_p$ is the variance in the positions of ions.}
\label{fig:intensity}
\end{center}
\end{figure}

To facilitate the further applicability of the observed coherent scattering and the related prospect experiments with improved axial optical access or even with different experimental platforms, we have analyzed the evolution of another important characteristics of multi-slit interference experiments: the scaling of the maximal intensity.
We have analysed its scaling within the voltage range where ion crystal remained stable in our trapping and cooling conditions. The maximum intensity has been evaluated after normalizing the light scattered by each ion to one. The observed dependence growth is expected to be naturally limited by the finite probability of the coherent scattering process and by the non-equidistant spacing between the neighbouring scatterers. The Fig.~\ref{fig:intensity} shows the measured normalized peak intensities plotted as a function of the ion number \emph{N} in our setup (green diamonds). The red dashed line shows the estimated case with fully incoherent scattering for the same ion numbers corresponding to the incoherent sum of scattered unity intensities. Importantly, the presented measured data show the intensity substantially higher than this incoherent case for all analysed cases up to 20 ions. Note the square root vertical axis scaling used for improving the readability. The upper bound of the scaling for emitters in harmonic trap would correspond to fully coherent scattering observable within the measured tip voltage ranges marked by orange circles. These are evaluated from the simulated ideal interference patterns by looking for the observable intensity maxima. As expected, the measured normalized peak intensities are much lower than in the case of fully coherent scattering. In addition, peak intensities for fully coherent case of scattering from emitters captured in a harmonic potential are lower than in the case of ideally equally spaced emitters due to the limited tuning range of practically achievable ion string lengths in our apparatus. On the other hand, the peak widths shown in the Fig.~\ref{fig:fwhm} are not affected by a portion of incoherently scattered light and their shape is given by the positions of scatterers. From experimental point of view relevant for further applicability of the presented results, it is worth noting that the idealized distribution of the perfectly equidistant scatterers is extremely hard or even impossible to achieve in most physical platforms. That is due the high sensitivity to even small asymmetry of the mutual distance between emitters or unbalance of their intensities. To allow for some rough comparison of realistic cases we have included the simulation of scattering from particles which have some random but fixed Gaussian uncertainty $\sigma_p$ in their position around the idealized center position corresponding to the equidistant case. The presented simulations correspond to the average from 100 random realizations. While few experimental platforms could with enormous effort reach the sub-wavelength deterministic equidistant positioning of single emitters and, at the same time, their sub-wavelength spatial extent, most of these realizations will be actually limited to the precision on the order of scattered light wavelength which will give a limit on the precision of detection of the particle position. The simulations corresponding to $\sigma_p=\lambda/2$ and $\sigma_p=2\lambda$ are shown by light blue and dark blue triangles, respectively. These results show that the position error of $\sigma_p=2\lambda$ would be almost equivalent to the presented non-equidistant case corresponding to harmonic potential within the considered number of scatterers. It is important to recall here that the presented non-equidistant case formed by the quadratic potential and the mutual Coulomb repulsion defines the relative position of the scatterers in a predictable fashion and with very small uncertainty. This is contrary to the equidistant case where scatterers have to be typically positioned independently by creation of many individual binding potentials. A feasible experimental way towards the utilization of the observed coherent scattering in the arrays of equidistant atomic scatterers could be represented by trapping of ions in optical lattices, where trapping potentials at multiples of trapping light wavelengths can be formed naturally~\cite{bylinskii2015tuning, schmidt2018optical, laupretre2019controlling}. These systems have already demonstrated the crucial control and tunability of two and three-dimensional ion crystal structures~\cite{laupretre2019controlling}.

\section{Conclusions}

The coherent scattering from trapped-ion crystals containing up to 53~ions correspond to the first unambiguous proof of collective coherent interaction of light with such large ensembles of individual quantum emitters. Similar observations have so far been limited to the domain of single or very few scatterers limited mostly by the their degree of control and by the geometry of interference arrangement~\cite{eschner2001light, wolf2016visibility, eichmann1993young, neuzner2016interference,slodicka2012interferometric}. The observed interference patterns have the potential to trigger substantial experimental progress in many fundamental and applied research directions which require collective coherent enhancement from many scatterers and, at the same time, high control over their external and internal degrees of freedom and the fundamental possibility of their addressing~\cite{lehmberg1970radiation, porras2008collective, skornia2001nonclassical, sutherland2016collective, scully2009collective, kauten2017obtaining,cabrillo1999creation, slodivcka2013atom, thiel2007generation, maser2010versatile, ficek2002entangled,vogel1985squeezing, devoe1996observation, jin2010squeezing, casabone2015enhanced, oppel2014directional, clemens2003collective, mewton2007radiative}. We anticipate the direct extension of the presented results to optical generation of large atomic Dicke states, which can be transformed using addressed qubit operations to a large spectrum of Dicke states with broad application prospects~\cite{kobayashi2014universal}. The conservative estimate of generation rate of 20~ion W~entangled state sharing single collective excitation would be on the order of few~Hz for the realistic overall detection probability of scattered photon of 0.1~\% and probability of double-excitation below 10\,\%~\cite{araneda2018interference,cabrillo1999creation}. This approach could substantially outperform conventional entanglement generation methods employing coulomb interaction~\cite{monz201114}. The addressed spin flips or phase shifts on individual ions prepared in a collective state of internal atomic levels could be utilized for efficient control of the direction and phase of the collective optical emission, which will allow for harvesting the unique possibilities of high fidelity internal state preparation in trapped ion systems for deterministic control and efficiency of the coupling between light and atoms in free space. The demonstrated light coherence will, together with provably nonclassical statistical properties of light emitted from large ionic crystals~\cite{obvsil2017nonclassical}, directly enable the realization of new controllable nonclassical light sources based on ion crystals.

\section*{Acknowledgments}

We acknowledge technological support from the group of Rainer Blatt including the contribution of Yves Colombe and Kirill Lakhmanskiy to the construction of the employed linear Paul trap. We thank D. H. Higginbottom for useful comments on the manuscript. This work has been supported by the grant No.~GB14-36681G of the Czech Science Foundation and Palack\'y University IGA-PrF-2019-010.

\section*{Appendix A: Illustration of individual coherent contributions}
\label{subsec:illustration}
\setcounter{equation}{0}
\setcounter{figure}{0}
\renewcommand{\theequation}{A\arabic{equation}}
\renewcommand{\thefigure}{A\arabic{figure}}

The ratio of the coherent to incoherent contribution of light scattered from ions is extremely important and can be inferred from the data shown in the Fig.~2-g) in the main part of the article. However, the evaluated visibility is a global figure of merit and it does not reflect the possible inhomogeneity of amplitude of coherent contributions from ions. An example of such situation could be caused by inhomogeneous intensity distribution of the scattered laser beam across the long ion strings.
To illustrate the dependence on the  coherent contribution of each individual ion of the interference signal we consider the measurement with 20~ions.  In this case, the measured interference is well described by the simple simulation where each scatterer contributes equally.

\begin{figure}[h!]
\begin{center}
\includegraphics[width=1.0\columnwidth]{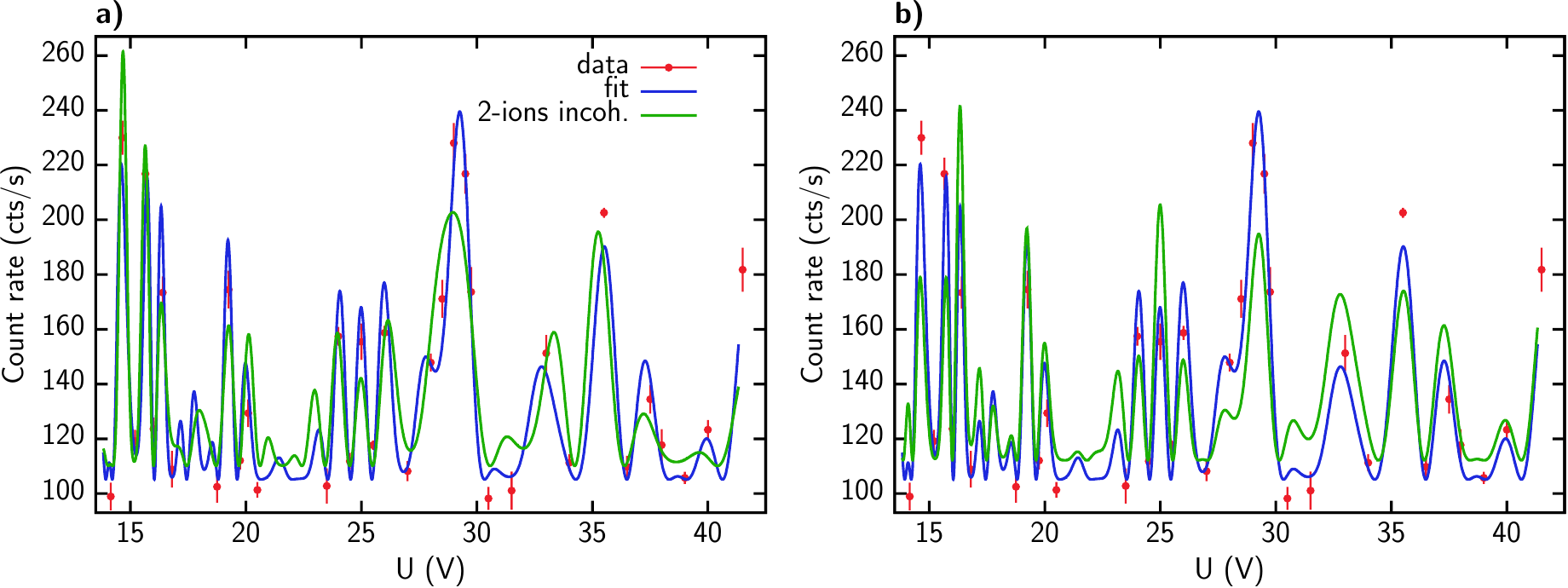}
\caption{Illustration of the impact of the relative change of the coherent contribution among ions in the string on the reproducibility of the measured phase interference patterns. The red points correspond to the measured data and the blue solid line is the best fit using the simulation where all 20~ions contribute with the same coherent contribution. The green solid line shows the best fit using the simulation where some of the ions were switched to fully incoherent scatterers. Graph a) illustrates the comparison of equal coherent contribution to the case where outermost two ions were switched to fully incoherent scatterers. In graph b), the comparison to simulation where the innermost two ions in the string are contributing only incoherently.}
\label{fig:false_fit}
\end{center}
\end{figure}

If we would consider that some of the ions scatter fully incoherently, the expected interference signal notoriously differs from the observed one.
For example, if the two outermost ions are set to emit fully incoherent light, they still contribute to the overall offset count rate, but do not contribute to the shape of the observable intensity pattern. Importantly, due to the re-scaling of the offset $I_{\rm incoh}$ and amplitude parameter $\kappa$ to obtain the best fit, this procedure is independent of the actual incoherent contribution. However, keeping the ions rather than removing them completely still allows us to simulate the more relevant case of ions being present but not contributing coherently. We note that complete removal of the ions from the string causes expectable radical shift and change of the whole interference pattern, because the relative positions of ions for given tip voltage would substantially differ. The simulation with two outermost ions switched to incoherent scaled to best fit the measured data is shown with a solid green line in Fig.~\ref{fig:false_fit}-a). The obvious negative impact on the ability to reproduce the measured pattern is most notable in the hight of constructive interference peaks. While some are enhanced, e.g., the first from the left, few other stay approximately similarly high, and some peaks are substantially suppressed, e.g., the peaks at 19~V and 29~V. The attempted best fit thus cannot anymore reproduce the data plausibly.
The situation is phenomenologically similar for the case where the two innermost ions emit fully incoherent light, see Fig.~\ref{fig:false_fit}-b). Switching of only single innermost or outermost ion to incoherent scatterer would show very similar behaviour to the above mentioned cases where two ions were symmetrically switched to incoherent and it would result in half of the observed effect of two ions in the simulated interference patterns.

\begin{figure}[b!]
\begin{center}
\includegraphics[width=1.0\columnwidth]{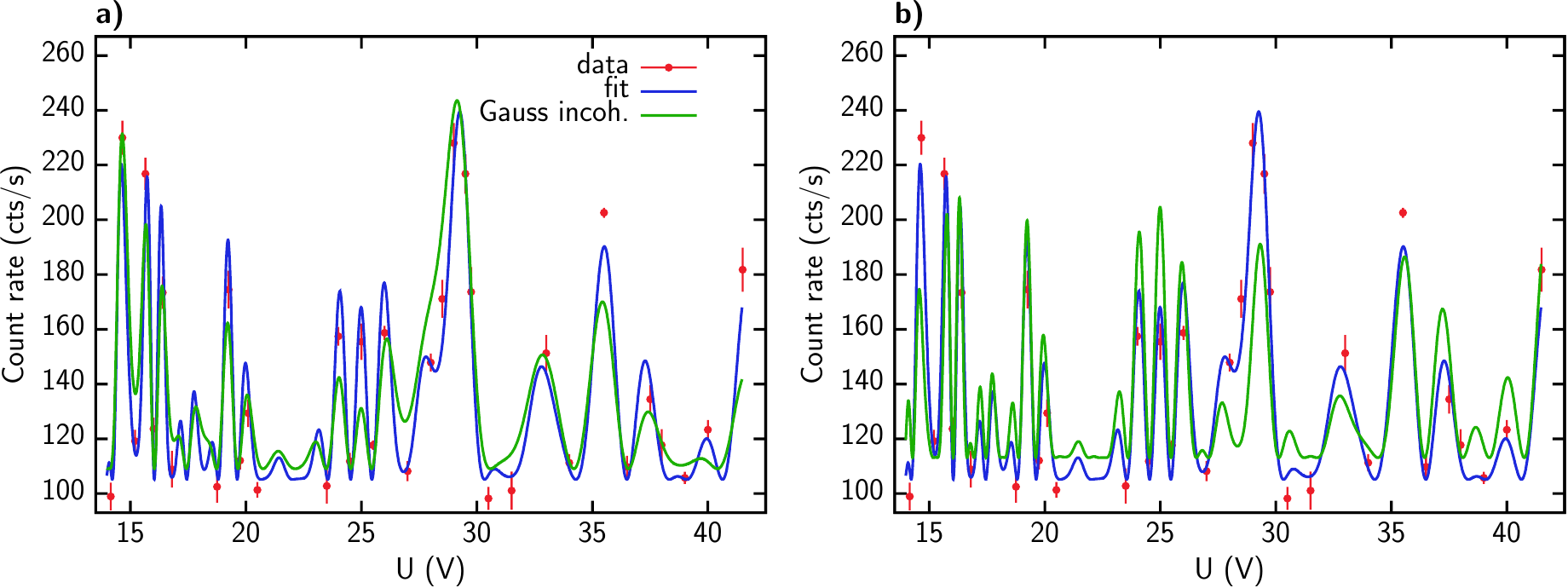}
\caption{Illustration of the impact of the relative change of the coherent contribution among ions in the string on the reproducibility of the measured phase interference patterns. The red points correspond to the measured data and the blue solid line is the best fit using the simulation where all 20~ions contribute with the same coherent contribution. The green solid line shows the best fit using the simulation where some of the ions were switched to partially incoherent scatterers. Graph a) illustrates the comparison of equal coherent contribution to the case of Gaussian modulation of the coherent contribution. The fully coherent scatterers are in the middle while at the each edge of the crystal the emitters scatter coherently only with the probability of $1/e$. In graph b), the coherent and incoherent parts of the gaussian modulation were switched, that corresponds to fully coherent scattering on the outermost ions in the string and coherent scattering with the probability of $1/e$ in center of the string.}
\label{fig:false_fit_gauss}
\end{center}
\end{figure}

Perhaps the most likely situation corresponds to the case where the fraction of the coherent scattering changes continuously with the distance from the center of the ion chain which coincides with the center of the exciting laser beam. We take this into an account by implementing the Gaussian distribution of the coherent to incoherent fraction of scattering processes as a function of the distance from center of the string into the simulation.
We consider two particular threshold cases. First case corresponds to the ions in the string center scattering fully coherently and the outermost ions are scattering with $1/e$ fraction of the coherent contribution, see Fig.~\ref{fig:false_fit_gauss}-a). The second case is the inverse of the first one, the outermost ions scatter fully coherently and the ions in the string center contribute with light that has $1/e$ coherent fraction. Similarly as in the case of switching some ions to fully incoherent, these simulations lead clearly to patterns which can not plausibly reconstruct the observed data.

Note that these illustrations are not a proof of the strictly equal coherent contribution from all ions in the string. However, they show that it is unlikely that the observed interference patterns can be explained without substantial relative contribution of any ion captured within the measured ion string. In addition, they directly point on the high sensitivity of the observed interference effect to the individual ion contributions which suggest the applicability of the presented scheme for partial or full switching of the scattered light in the given observation direction.

\section*{References}
\bibliographystyle{iopart-num}

\providecommand{\newblock}{}

\end{document}